\documentclass[sigchi]{acmart}

\usepackage{booktabs} 
\usepackage{todonotes}
\usepackage{multirow}
\usepackage{placeins}
\usepackage{siunitx}

\begin{document}
\title{In a Silent Way}
\subtitle{Communication Between AI and Improvising Musicians Beyond Sound}

\copyrightyear{2019} 
\acmYear{2019} 
\setcopyright{acmlicensed}
\acmConference[CHI 2019]{CHI Conference on Human Factors in Computing Systems Proceedings}{May 4--9, 2019}{Glasgow, Scotland Uk}
\acmBooktitle{CHI Conference on Human Factors in Computing Systems Proceedings (CHI 2019), May 4--9, 2019, Glasgow, Scotland UK}
\acmPrice{15.00}
\acmDOI{10.1145/3290605.3300268}
\acmISBN{978-1-4503-5970-2/19/05}
\settopmatter{printacmref=true}
\fancyhead{}

\author{Jon McCormack}
\orcid{0001-6328-5064}
\affiliation{%
  \institution{SensiLab, Monash University}
  \city{Caulfield East}
  \country{Australia}
}
\email{Jon.McCormack@monash.edu}

\author{Toby Gifford}
\affiliation{%
  \institution{SensiLab, Monash University}
  \city{Caulfield East}
  \country{Australia}
}
\email{Toby.Gifford@monash.edu}

\author{Patrick Hutchings}

\orcid{0001-8680-6969}
\affiliation{%
  \institution{SensiLab, Monash University}
  \city{Caulfield East}
  \country{Australia}
}
\email{Patrick.Hutchings@monash.edu}

\author{Maria Teresa Llano Rodriguez}
\affiliation{%
  \institution{Goldsmiths, University of London}
  \city{London}
  \country{United Kingdom}
}
\email{m.llano@gold.ac.uk}

\author{Matthew Yee-King}
\affiliation{%
  \institution{Goldsmiths, University of London}
  \city{London}
  \country{United Kingdom}
}
\email{m.yee-king@gold.ac.uk}

\author{Mark d'Inverno}
\affiliation{%
  \institution{Goldsmiths, University of London}
  \city{London}
  \country{United Kingdom}
}
\email{dinverno@gold.ac.uk}

\renewcommand{\shortauthors}{J. McCormack et al.}

\begin{abstract}
Collaboration is built on trust, and establishing trust with a creative Artificial Intelligence is difficult when the decision process or internal state driving its behaviour isn't exposed. When human musicians improvise together, a number of extra-musical cues are used to augment musical communication and expose mental or emotional states which affect musical decisions and the effectiveness of the collaboration. We developed a collaborative improvising AI drummer that communicates its confidence through an emoticon-based visualisation.  The AI was trained on musical performance data, as well as real-time skin conductance, of musicians improvising with professional drummers, exposing both musical and extra-musical cues to inform its generative process. Uni- and bi-directional extra-musical communication with real and false values were tested by experienced improvising musicians.  Each condition was evaluated using the FSS-2 questionnaire, as a proxy for musical engagement. The results show a positive correlation between extra-musical communication of machine internal state and human musical engagement.
\end{abstract}

%
%



\begin{teaserfigure}
  \centering
   \includegraphics[width=0.28\textwidth]{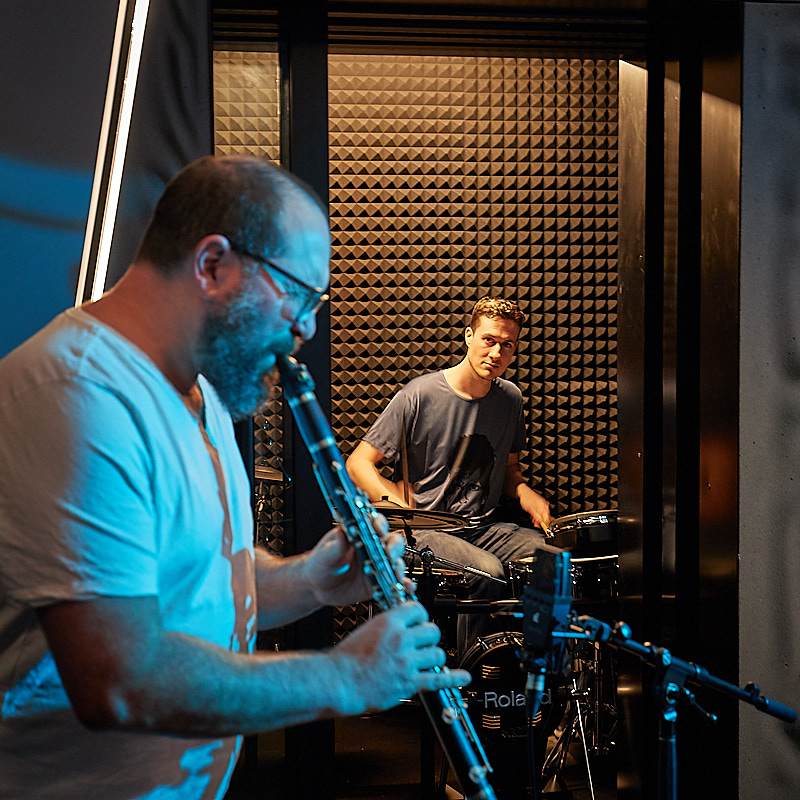} \hspace{40mm}
   \includegraphics[width=0.28\textwidth]{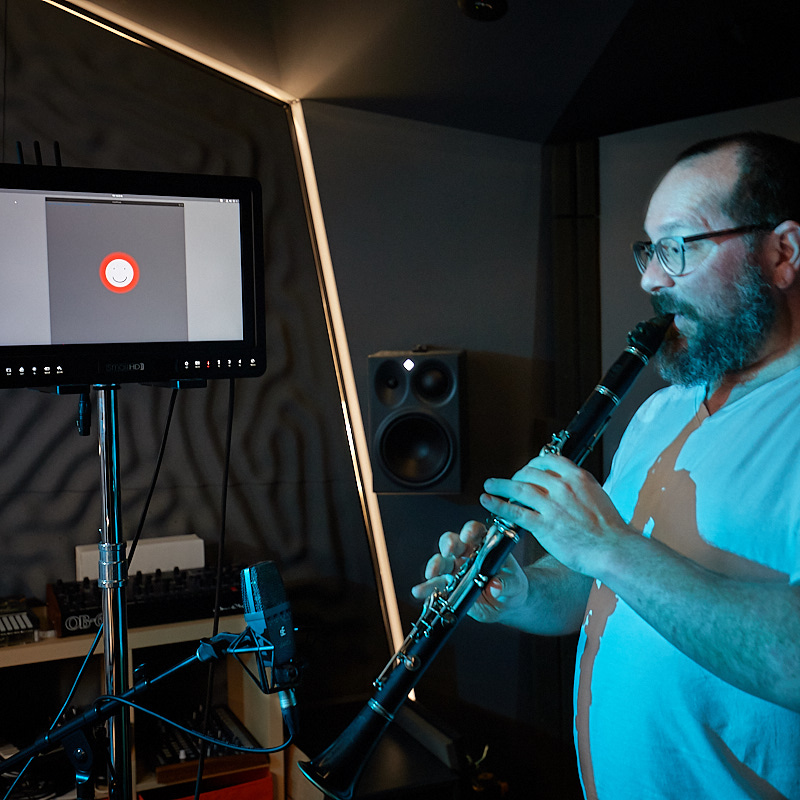}
   \caption{Training with percussionist and instrumentalist improvisation (left); performance with the AI Improviser (right)}
   \label{f:improv}
\end{teaserfigure}

\maketitle

\section{Introduction}
In this paper we investigate the benefits of extra-musical interaction in real time music improvisation and co-creation with an artificially intelligent creative partner. Recent advances in Artificial Intelligence (AI) techniques, coupled with increasingly powerful computer resources, make it feasible to engage in artistic collaborations with a machine intelligence, because it exhibits degrees of creative agency \cite{BownM09} and autonomy \cite[Chapter 9]{Boden2010} which traditional tools or instruments -- either digital or analogue -- do not possess \cite{dinvernoMcCormack2015,McCormackdInverno2016}. 

Interactions between musical improvisers largely occur through the music itself, although visual and other extra-verbal cues play an important role~\cite{Sawyer2003c}. We explore the role of extra-musical cues and feedback between human and AI performers to achieve better performance outcomes with a longer term goal of establishing trust between human and machine -- widely acknowledged as an important factor in successful human group improvisations~\cite{Waterman2015,seddon_modes_2005}. We are also interested in understanding how human-AI improvisations with and without extra-musical cues are perceived by audiences; our research also evaluated this aspect of resultant performances between human and AI.   

\subsection{Improvisation and Extra-musical Communication}
Live performance and improvisation are amongst the most challenging creative activities undertaken by humans. To do them successfully requires a great deal of proficiency and virtuosity which typically takes many years of practice and experience before one can claim anything close to mastery \cite{Eri93,Gla08,pachet2012}. In musical improvisation the ability to communicate allows participants to be aware of and understand the behaviour and intentions of those involved. Working within an improvised performance setting without some level of understanding of what others are experiencing makes it very difficult for performers to create settings that inspire each other and take the performance forward.

When it comes to improvising with a non-human performer, important traditional cues and indicators may be missing: body language and movement, eye contact, visual cues, etc., do not exist. For the performer this means that all of the AI's intentions must be inferred through the musical output. This makes it more difficult to build trust and, therefore, to take risk during a performance; both trust and risk-taking are widely considered important aspects of successful group collaboration and teamwork~\cite{Waterman2015,jana_conceptualizing_2018}.

\subsection{Musical Engagement and Flow}
Theories of human musical engagement typically posit a phenomenological state known as \textit{flow} \cite{csikszentmihalyi1990flow} as both an underlying driver and fundamental metric of engagement \cite{de_manzano_psychophysiology_2010,wrigley2013flow,Jackson2002}. A flow state is characterised by a number of cognitive, affective and psychophysiological indicators. Cognitive factors include a sense of effortless control and complete focus \cite{EJOP1370}. Affective factors include loss of self-consciousness and high intrinsic motivation \cite{landhausser2012flow}. Psychophsyiological markers of flow include salivary cortisol, blood pressure, and heart rate variability \cite{de_manzano_psychophysiology_2010} as well as skin conductance \cite{nakamura2016hypo}.

Flow is theorised to be an important aspect of successful group improvisations \cite{Sawyer2003c,hytonen2016experiencing,Robb:2015aa} and has also been discussed as an engagement metric for human-machine creative partnerships \cite{pachet200619,Gifford:2017aa}.

\subsection{Biometrics During Music Performance}
Biometrics can provide real-time quantification of human psychophysiological state with relatively minimal distraction, and thus may be useful for communicating human mental state to an AI during live performance. Given the importance of flow states to group musical improvisation, the biometric markers discussed above for flow suggest themselves as relevant variables to be exposed. In this experiment we selected skin conductance as an indicator of human internal state.

Skin conductance (SC) has been studied in the context of music performance by Dean \& Bailes who note that ``SC measures are frequently interpreted as an index of not only emotional response ... but also task effort and attention'' \cite{dean2015sc}. They find that time-series analysis of real-time SC can predict musically salient features of an improvisation. The relationship between SC and human internal state is, however, not straightforward, possibly due to the ``inherent complexity of flow experiences'' \cite{nakamura2016hypo}. Of interest in this study then was whether or not a machine learning system could extract useful information from a real-time SC measure.

\subsection{Aims and Contributions}
This study tests two related hypotheses in human-machine musical improvisation. We hypothesise that extra-musical communication of `internal state':
\begin{enumerate}
\item \textbf{\textit{of} the human musician \textit{to} the machine} can enhance the machine's capacity to generate appropriate and complementary improvised output;
\item \textbf{\textit{of} the machine \textit{to} the human} can facilitate more engaging human-machine musical interactions.
\end{enumerate}
Our study employs a factorial design, where a series of human-machine musical improvisations are evaluated under combinations of conditions: with/without human-to-machine extra-musical communication, and with/without machine-to-human extra-musical communication.

Results of both detailed evaluations from improvising musicians using the system and a listening study of 100 external observers conducted on-line support the use of machine-to-human extra-musical communication in the form of machine confidence visualisations.  When truthful, such visualisations were found to produce a higher reported flow state \cite{csikszentmihalyi1996flow} in performers than reported when using deceptive or absent visualisations, on average.  Significantly more listeners perceived a better musical balance between machine- and human-performed instrumental parts in recordings made when truthful, rather than deceptive, confidence visualisations were used during recording.

Human-to-machine extra-musical communication, in the form of skin conductance measurements, was not found to have a significant effect on either training the AI system or on the flow state of performers, probably due to confounding noise from muscle movements.

\section{Related Work}
Many different factors have been identified as significant to increasing trust in human-machine collaborations, including reliability, predictability, utility, provability, transparency and explainability \cite{Muir1987,Lee1994,Lee2004}. We are particularly interested in how revealing intrinsic aspects of the workings and purpose of both humans and machines can influence trust and promote engagement. In other words, we are interested in revealing the state of a human-machine collaboration in a way that helps both parties understand the interaction taking place. Sawyer \cite{sawyer:2000} suggests that improvisational creativity in a collaborative performance is achieved through ephemeral signs; thus, the way an AI improviser communicates must be simple and precise,
yet the communication itself needs to be meaningful 
so that the new information helps progress the performance. 

\subsection{Extra-musical Communication}
Research addressing issues in extra-musical communication between human and AI improvisers is currently in its infancy. 

Weinberg, Hoffman and Bretan developed a series of expressive robotic improvising musicians~\cite{Hoffman2011,bretan2017towards}, most notably their improvising jazz marimba robot \textit{Shimon}. The physical embodiment of \textit{Shimon} is an important aspect of their research, both in terms of its extra-musical communication through movement, which affords human-machine temporal co-ordination through anticipation, and its visually animated appearance, lending it the impression of musical personality~\cite{BRETAN20151}. Bretan suggests that an obvious next step for this area of research is to incorporate ``social cues'' to convey musical emotion and ``lead to more convincing performances by the robot in which the system looks truly expressive''~\cite{bretan2017towards}.

Ravimukar et al.~\cite{ravikumar_notational_2017} describe a research proposal centred on the question: ``will the addition of two-way extra-musical notational communication enhance the human's experience of coordinating musical transitions with AI music partners?'' which bears similarity to our study, although focusing on notational communication for temporal anticipation of musical changes. Their study is in progress, and does not appear to have reported results at the time of writing.

Skin conductance (also known as \textit{galvanic skin response}) has been used to train generative music systems with the goal of producing controlled, affective output.  Kim and Andr{\'e} \cite{kim2004composing} utilised galvanic skin response, along with electrocardiogram, electromyogram and respiration data, as input for a generative music system that used genetic algorithms.  While the input is provided by a human listener they are not presented as a collaborative partner in producing music together.  Hamilton \cite{hamilton2006bioinformatic} developed a real-time composition system that records galvanic skin response of a human performer and uses this measure as input for a software composition system that generates notes on a score for the performer to play in real-time.

\subsection{AI generative systems and confidence}

\begin{figure*}[t!]
  
  \centering
    \includegraphics[width=0.67
    \textwidth]{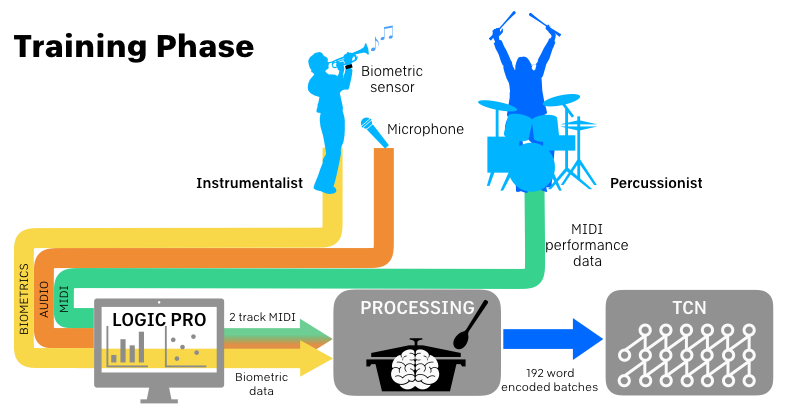}
    \caption{Training the machine improviser with improvisations on the electronic drum-kit and melodic instrument.}
    \label{training}
\end{figure*}

Neural network-based systems have been utilised for symbolic music generation for decades~\cite{eck2002first} and have seen a resurgence along with the deep learning boom of the last five years.  Increased model complexity has seen improvements in effective memory, expressive range and consistency of generated outputs, making neural networks a good candidate for collaborative AI music systems.  

Recurrent Neural Networks (RNNs) have been the predominant neural network architecture for music generation tasks~\cite{briot2017deep}, including drum sequence generators~\cite{makris2018deepdrum,hutchings2017talking}.  Recent research has shown however that Temporal Convolutional Network (TCN) models~\cite{rene2017temporal} can perform just as well or better in the analysis of sequential data~\cite{bai2018empirical}.  TCN models also have a greater number of parallel pathways that allow for models to be trained significantly faster on GPUs.  

In sequential data models, a softmax layer can be used to predict upcoming events and post-assessed in terms of an entropy metric, or an accumulated improbability score. Softmax layers output real values between 0 and 1 such that all values in a specified dimension sum to unity, and as such can be used to represent a probability distribution.

When creating a generative system from a predictive model, probability distributions can be sampled to generate an output.  By doing so, the distribution is collapsed into a single choice and the majority of data output from the network itself is discarded.

\section{System Implementation}
We implemented a machine improviser and studied the experience of musical co-creation with it. The machine improviser is software utilising a Temporal Convolutional Neural Network (TCN)~\cite{rene2017temporal} -- a machine learning system to drive real-time algorithmic generation of percussive accompaniment in a musical duo. In particular, the study probes the efficacy of bi-directional communication of `internal state' via extra-musical communicative channels between improvisers.

The research design included data gathering from human musicians in two distinct stages: (i) training of a TCN software drummer based on human-human improvisational duets (Figure \ref{f:improv}, left and Figure \ref{training}), and (ii) evaluation of the experience of co-creating with the trained TCN under various conditions of extra-musical communication (Figure \ref{f:improv}, right and Figure \ref{f:evaluating}).

The network was trained on 3 hours of human duet improvisation by experienced improvising musicians.  Duets were performed between an instrumentalist playing melody on a monophonic instrument (variously saxophone and clarinet) and a drummer playing an electronic drum-kit. 

\subsection{Musical performance data collection}

The drum-kit, a Roland TD-50, is an electronic kit designed to closely emulate the sound and feel of a standard acoustic drum-kit. It records and transmits performance data in extended precision (14 bit) MIDI format, whilst also synthesising emulated acoustic sounds via physical modelling~\cite{webb2014parallel}.

Musical input from the human instrumentalist (clarinet, saxophone) was recorded as audio and algorithmically transcribed into MIDI format using Logic Pro's \emph{Flex Audio}\footnote{\url{https://ask.audio/articles/logic-pro-x-tutorial-flex-pitch}} feature. 

The training data was gathered over four sessions, with two different drummers and two instrumentalists in all combinations. The sessions took place in a recording studio, and utilised a `click-track' to define the underlying tempo and beat.

Each session had 9 exercises comprising combinations of 3 musical styles and 3 performance techniques in a factorial design. The styles were (i) Swing, (ii) Funk and (iii) Rock, all in common (4/4) metre, at a fixed tempo of 120 bpm. The performance techniques were (i) melodic lead with percussive accompaniment, (ii) trading groups of four measures between performers, and (iii) trading groups of two measures.

The use of a click-track facilitated symbolic transcription of the recorded duet improvisations as beat-relative note events, quantised to 12 time steps per beat, allowing triple and duple subdivisions down to the resolution of semi-quaver triplets. After symbolic transcription, the musical data was tokenised for feeding into the TCN model.  Each four-beat measure of the performance then comprised a 48 token string.  

Tokens conveyed if a note was started (\textit{H}) or sustained (\textit{s}) by the melodic instrumentalist and the onset velocity, quantised to four bands (\textit{p},\textit{mp},\textit{mf} and \textit{f}).  These tokens were concatenated with tokens for each drum hit encoded as midi pitch values and the same velocity metric to form longer tokens.  This example phrase segment shows how sequences were tokenised with this method: \textit{38mp o 36mf|38mf|44mf o 38mp o o 36mp|38mp Hmp s s|38mp Hmp s|38mp s s s}.  A total of 1639 unique tokens were required to encode the training data.  Of these, 1188 tokens were used less than 20 times and were replaced with the silent token `o', leaving 451 tokens used in the corpus.

\begin{figure}[b!]   
  \centering
    \includegraphics[width=0.48
    \textwidth]{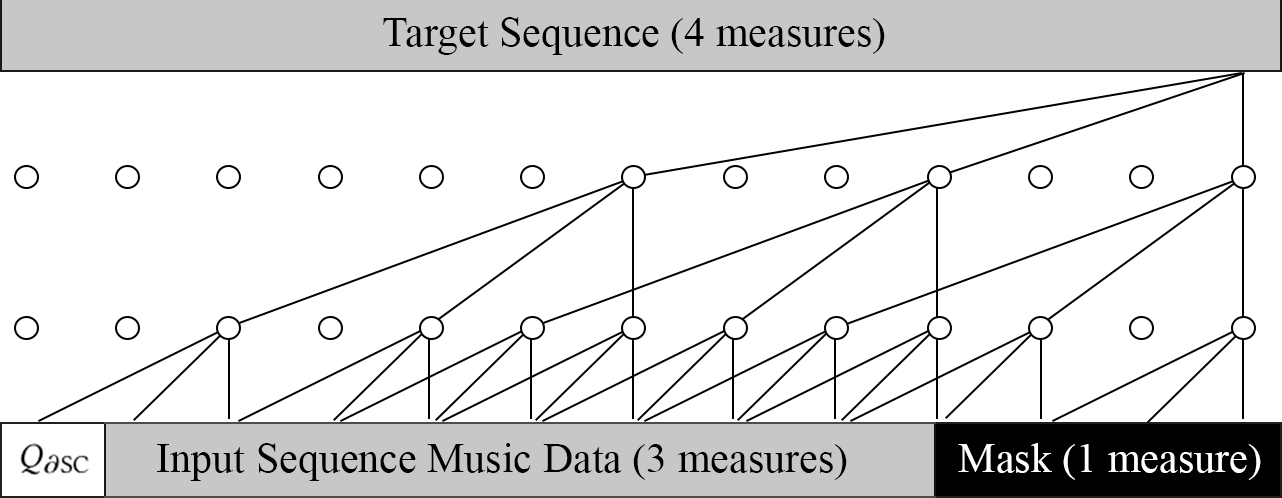}
    \caption{A simplified representation of the TCN model, showing how the final token of the target sequence is predicted with the use of skin conductance and quantised music data.}
    \label{TCN}
\end{figure}

\subsection{Biometric data collection}
In addition to translating the musical performance data as input for the TCN model, the instrumentalist wore an Empatica E4 biometric wristband\footnote{\url{https://www.empatica.com/research/e4/}}, which recorded real-time skin conductance (SC). Following Dean and Bailes~\cite{dean2015sc} we utilised change in skin conductance ($\operatorname{\partial SC}$) as a real-time parameter containing information about the human musician's mental state and cognitive music processing.

The Empatica E4 wristband reports skin conductance in microSiemens at a sampling rate of 4 Hz. Baseline skin conductivity varies between people and its absolute level as measured by the wristband depends on how tightly the band is fitted. As such we used a relative measure of change in skin conductance defined by
$$\operatorname{\partial SC}_t = \frac{\operatorname{SC}_t - \operatorname{SC}_{t-1}}{\sigma_{\operatorname{SC}}}$$
where $\sigma_{\operatorname{SC}}$ is the standard deviation of the skin conductance over 2 minutes prior to commencing the improvisation session. A further 1 minute period (minimum) waiting period was used from the time the wristband was put on before commencing baseline calibration, to exclude the large changes related to fitting the wristband.
The real-time value of $\operatorname{\partial SC}$ was quantised to 3 discrete levels: 
\[ Q_{\operatorname{\partial SC}} = \begin{cases} 
      \operatorname{High} & \operatorname{\partial SC} \geq 1 \\
      \operatorname{Med}  & -1 < \operatorname{\partial SC} < 1 \\
      \operatorname{Low}  & \operatorname{\partial SC} \leq -1
   \end{cases}
\]

Across the performance sessions High, Medium and Low values were recorded for 15\%, 65\% and 20\% of the measurements respectively. The quantised signal was then sampled at the start of each musical measure and communicated to the machine improviser via an OSC~\cite{wright2005open} request and response.

For communication to the human musician of the machine improviser's internal state, we expose a proxy index of the evolving `confidence' reported by the neural net regarding its musical decisions, calculated from its internal distribution of probabilities. We elaborate on this later in this section.

\subsection{Training the machine improviser}

\begin{figure*}[t!]
  
  \centering
    \includegraphics[width=0.7
    \textwidth]{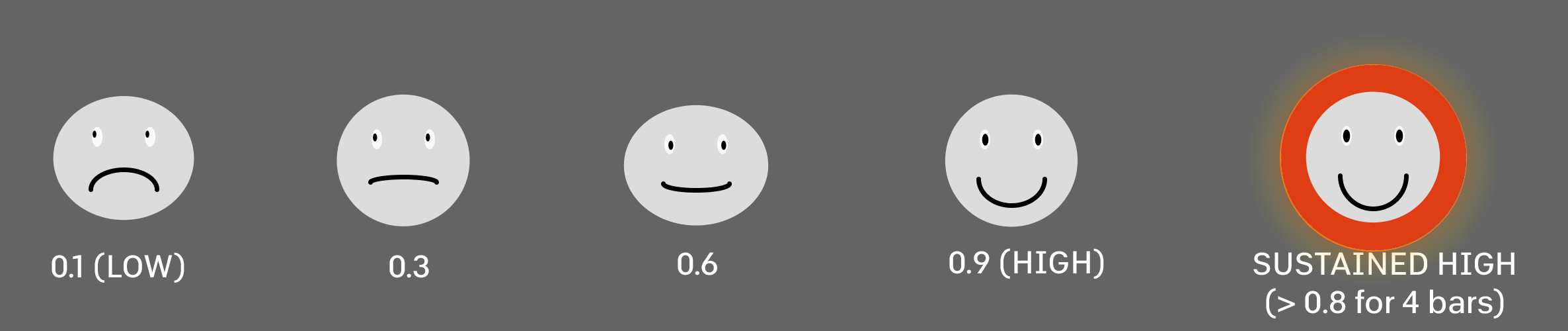}
    \caption{Still images of the visualiser at different levels of confidence ranging from 0.1 (low confidence) to 0.9 (highly confident). The expression changes continuously in response to the confidence of the machine improviser in addition to nodding in time with the current tempo. Sustained high levels of confidence result in a pulsating glow behind the emoticon in time with the beat.}
    \label{f:ConfidenceVis}
\end{figure*}

The machine improviser is a software system comprising a TCN that generates performance data for a drum synthesiser to perform over the next measure of the performance, run consecutively every measure.  The TCN model~\cite{bai2018empirical} was trained to predict combinations of notes played in any given measure of performance based on musical and biometric data from the three measures directly preceding it.  By sampling from these predictions, a generative system was produced.

The improvisation sessions used for training data provided a set of 4195 sequences of four measures which was divided into training (76\%), validation (12\%) and test sets (12\%), which provided a natural split in the data.

During training, the network used four measures of performance data for input and target sequences.  The fourth measure of the input sequence was masked to prevent the system from using any time step from the fourth measure to generate predictions.  This allows the whole fourth measure to be predicted (and generated) at once, in parallel.

To facilitate an effective memory to cover the full 192 time steps in four measures, dilations were used through seven convolution layers. With a kernel size of 3 and 192 units per hidden layer, a receptive field of 257 time steps was achieved.  The model used for training can be found at \url{https://github.com/patHutchings/TCN/tree/Machine-Improviser}

\begin{figure*}[htbp]
  
  \centering
    \includegraphics[width=0.67
    \textwidth]{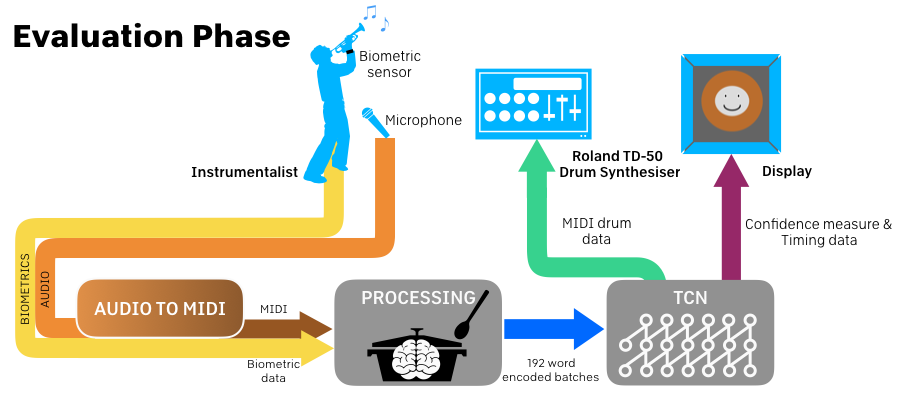}
    \caption{Evaluating the machine improviser with instrumentalists.}
    \label{f:evaluating}
\end{figure*}

Two unique data preprocessing features were implemented for training and inference with the TCN model.  In sequence-to-sequence tasks, where the next token in a sequence is predicted, the input sequence is typically offset from the target sequence by inserting a `<start>' token or similar at the beginning of the input sequence.  $Q_{\operatorname{\partial SC}}$ from the last measure is used as the start token, making it visible at every step for training and inference (Figure \ref{TCN}).

An extensive hyper-parameter search was conducted to reduce model complexity without loss of predictive accuracy, as measured by per-token perplexity on a validation set of 500 four measure strings.  Perplexity represents the average number of most probable tokens to appear next in the sequence.  Complexity was reduced by using a small token embedding of only 20 dimensions and limiting effective memory to four measures.  The small model size allows inference of a measure of drum performance within 5ms on a consumer level 1080ti graphics card: less than a single time step in the quantised performance.

The silent `o' token appeared an order of magnitude more frequently than any other token in the training data and the TCN quickly learned that predicting entirely silent sequences would still return a small loss.  To counteract this effect a weighting of 0.1 was applied to `o' token contributions to loss calculations.

A perplexity of 6.70 was achieved on the validation set after 69 epochs of training with batches of 16 sequences.  This means the system finds an average of 6.70 tokens that are most likely to occur next at any point in the sequences, from the 451 tokens it knows.  A perplexity of 6.68 was achieved when $Q_{\operatorname{\partial SC}}$ was replaced with a standard '<start>' token, indicating that the inclusion of skin conductance data did not assist or impede the training of the network. The model with biometrics was used for the machine improviser and a final test set evaluation was performed, producing a perplexity of 7.34.

A softmax layer was used as the final layer of the TCN, so that outputs could be used as probability distributions for sampling in the generative system. The probability of any of the 451 tokens in the corpus dictionary being played at any timestep in the upcoming measure is used to make a weighted selection.  The machine improviser uses only tokens observed in the training data.  With this dictionary the system has \num{2.5e127} possible outputs at each measure. Tokens are decoded to MIDI messages and sent to the drum synthesiser for performance.

\subsection{Visualising Confidence}

The AI's confidence was conveyed visually using an emoticon-style face (Figure \ref{f:ConfidenceVis}) drawn with simple vector graphics, that bounces in time with the music with a refresh rate of 60hz.  Confidence values were updated every 0.5 seconds and reflected on the display within 5ms.  When system confidence is low, the face frowns and eyes shift in different directions, avoiding eye contact with the viewer.  When confidence is high, the face smiles, its eyes widen, and it maintains eye contact with the viewer.  Sustained high states produce a radiating glow behind the face that pulsates in time with the beat.

The use of a simplified, iconic representation of a face was chosen to minimise additional cognitive load on the musician, mimicking facial expressions at a high level of abstraction in ways similar to how human performers behave when they lack confidence in performance.  Iconic facial expressions are easy to process, but also avoid over-anthropomorphising the AI, which may potentially lead to false assumptions regarding the level of cognition, emotion, and behavioural characteristics if a more realistic or nuanced human face were used, for example.

\section{Performer Evaluation}

After developing and training the machine improviser, we then evaluated it musically through trials with improvising musicians. We describe the methodology and results below.

\subsection{Method}

We recruited seven experienced instrumentalists who each engaged in four improvised sessions for a total of 28 trials.  Recruited participants were given a ~\$15 department store voucher for their time.

Three female and four male instrumentalists aged between 18 and 56 took part in the evaluation.  While all had experience with improvisation, different approaches to improvisation were reflected in the range of instruments (saxophone, clarinet, vocals and electronic keyboard), styles of familiarity (jazz, rock, funk, experimental, folk and classical) and roles (professional musicians, serious amateurs and skilled hobbyists).  

In each session the instrumentalist improvised with the machine improviser for 3 minutes and then self-reported on the experience using the FSS-2 SHORT scale \cite{csikszentmihalyi1996flow}.  The FSS-2 SHORT scale comprises 9 items (i.e. questions) intended to capture the 9 dimensions of flow as described by Csikszentmihalyi \cite{csikszentmihalyi1996flow}. 

Application of the FSS-2 suite of scales to musical performance has found that these items have differing correlation with other external flow metrics, and so a subset of the items may be more appropriate for measuring flow in musical contexts. Wrigley and Emerson \cite{wrigley2013flow} found that the ``subscales of Sense of Control, Autotelic Experience, and Challenge-Skill Balance showed the strongest associations and explained the most variance'' in live music performance.

Evaluations took place in the same recording studio used for data collection, with the addition of a portable screen, positioned at standing height, which displayed the confidence visualisation (see Figure \ref{training}, right and Figure \ref{f:evaluating}).

We developed three different confidence visualisation conditions to differentiate effects driven by visualisation design choices from those driven by the content it is being used to communicate. Confidence was (i) truthfully communicated by the visualisation system, (ii) inverted to create a deceptive visualisation and (iii) not communicated by removing all facial features of the emoticon to create an `absent' confidence visualisation.  Inversion was used for the deceptive condition as an `absent' condition was tested for and inverted patterns provided changes between the communication states with the same transition dynamics.

Instrumentalists were given a short verbal brief to communicate that the visualisations were communicating a metric of confidence of the improvising machine in predicting what to play next.  It was emphasised that the visualisation is not a judgement on the instrumentalist's playing or overall quality of the musical performance.  The `intelligence' of the improvising machine was described as generating musical improvisations by repeating patterns learned to be useful in accurately predicting what might happen next in improvised duets we recorded.

\begin{table}[b]
\caption{Test conditions with truthful (T) and deceptive (D) biometric data for each visualisation type. Random selections represented with (T/D).}
\label{tab:conditions}

\begin{tabular}{@{}llll@{}}
\toprule
          & \multicolumn{3}{l}{Confidence Visualisation}                   \\ \cmidrule(l){2-4} 
          & Truthful          & Deceptive         & Absent                 \\ \cmidrule(l){2-4} 
Condition & Bio T/D & Bio T/D & Bio T and Bio D \\ \bottomrule 
\end{tabular}

\end{table}

Four of six possible conditions were tested with randomised order for each instrumentalist (see Table \ref{tab:conditions}). Priority was placed on having longer sessions of improvisation without fatiguing instrumentalists physically and creatively.

\subsection{Results and Discussion}

We found that the \textbf{visualisation of machine confidence} noticeably \textbf{affected the tendency} of the instrumentalist \textbf{to achieve flow.} The biometric communication via the instrumentalist's skin conductance did not make any discernible difference to the experience of improvising with the system. These results emerged from comparison of an aggregate flow measure derived from the 9 survey items in the FSS-2 responses, compared between conditions across participants.

We performed a Principal Components Analysis on the FSS-2 responses (comprising 4 sessions x 7 participants = 28 responses for each of the 9 questions). The first principal component was consistent with the findings of Wrigley and Emerson in having substantial loadings for Sense of Control, Autotelic Experience, and Challenge-Skill Balance, and negligible loadings for  Clarity of Goals and Transformation of Time. As such, we utilised an aggregate index of flow comprising the average of the questions relating to Sense of Control, Autotelic Experience, and Challenge-Skill Balance. For completeness we additionally ran all the analyses with an aggregate index constructed using the numeric weights contained in the first principal component of our data, which did not qualitatively change any of our conclusions.

The visualisation condition had a measurable impact on the aggregate flow index. The Truthful condition was more flow-inducing than the Absent condition, which in turn was more flow-inducing than the Deceptive condition. For 4 of the 7 participants (P1, P2, P3, P7) this relationship was monotonic across the three conditions, and for 5 of the 7 participants (P1, P2, P3, P6, P7) the Truthful condition showed equal or better flow induction than the Deceptive condition. Of the remaining 2 participants, 1 (P4) showed an inversely monotonic relationship, and the other (P5) strongly preferred either Truthful or Deceptive visualisation over Absent visualisation in terms of their flow index.

\begin{table}
\caption{flow index \textit{vs.} visualisation condition}
\label{tab:viscondflow}
  \begin{tabular}{cccc}
    \toprule
    Participant&Deceptive&Absent&Truthful\\
    \midrule
    1 & 3.67 & 4.33 & 4.33\\
    2 & 3.67 & 4.17 & 4.33\\
    3 & 3.33 & 4.17 & 4.33\\
    4 & 4.33 & 4.17 & 3.67\\
    5 & 4.00 & 2.83 & 3.67\\
    6 & 4.00 & 3.16 & 4.00\\
    7 & 2.00 & 3.33 & 3.67\\
    \midrule
    \textbf{mean} & \textbf{3.57} & \textbf{3.74} & \textbf{4.00}\\
    s. d. & 0.76 & 0.61 & 0.33\\
  \bottomrule

\end{tabular}

\end{table}

In order to assess the significance of these trends we performed matched-pair t-tests on each of the three between-condition combinations of the machine confidence visualisation: Absent \textit{vs.} Deceptive, Truthful \textit{vs.} Absent, and Truthful \textit{vs.} Deceptive. The number of instrumentalists in the study is small (N = 7). Student's t-test was originally developed for statistical inference from small samples \cite{ziliak2008guinnessometrics}, and Cummings \cite{CummingGeoff2012Utns} recommends reporting of 95\% confidence intervals derived from the t-distribution and the sample standard deviation for sample sizes between 5 and 30. Whilst some researchers promote non-parametric tests and much larger sample sizes for Likert scale analysis in HCI \cite{robertson_likert-type_2012}, Norman argues to the contrary that ``parametric statistics can be used with Likert data, with small sample sizes, with unequal variances, and with non-normal distributions, with no fear of coming to the `wrong conclusion' '' \cite{norman_likert_2010}, and recommends paired t-tests for comparing conditions via Likert scales for sample sizes of at least 5. 

The average effects of the machine confidence visualisation conditions on the flow index are summarised in table \ref{tab:effectvisflow}. The Truthful visualisation condition was on average more flow-inducing than the Deceptive condition, with an effect size of approximately 1/2 out of a scale of 5, significant at the 99\% level, and the Deceptive condition averaged 1/4 point lower than the Absent condition, significant at 95\%.
\begin{table}
\caption{effect of visualisation}
\label{tab:effectvisflow} 

  \begin{tabular}{rccc}
    \toprule
         & Absent & Truthful & Truthful \\
         & \textit{vs.} & \textit{vs.} & \textit{vs.} \\
         & Deceptive & Absent & Deceptive \\
         
    \midrule
    upper 95\% & 0.49 & 0.62 & 0.83 \\
    \textbf{mean}  & \textbf{0.26*} & \textbf{0.16} & \textbf{0.42**} \\
    lower 95\% & 0.03 & -0.28 & 0.03 \\
    \bottomrule
   
\end{tabular}

\end{table}

The statistical tests described above allow us to make inferences about underlying effects in the presence of random noise, such as the variable behaviour of the algorithmic improviser. They are not, however, designed to allow generalisations to people beyond the study participants. In any research, the only reliable way to make quantitative generalisations to people beyond the study group is to recruit participants by randomly sampling from the entire population of interest. However, qualitative judgements regarding probable transferability of study results can often be argued from the diversity of the study participants \cite{flyvbjerg_five_2006}. In our study we did not have any particular population in mind, though we hope the results may be broadly indicative of the type and diversity of reactions that experienced improvising musicians would have had if included. We employed convenience sampling in recruiting instrumentalists to evaluate the system, and even this small group had quite diverse range of approaches and responses. As discussed in \S \ref{subsec:external_listener_evaluation} the instrumentalists covered a broad range of musical styles, lending some confidence that our results may have relevance outside of the study group.

\section{Listener Evaluation}
The experience study showed that the participants' tendency to achieve flow was enhanced by extra-musical communication of machine confidence. But what about the musical output? To see if effects observed through changes in the confidence visualisation conditions were perceivable only to a performer, or were also noticeable by external listeners, we conducted an additional on-line listener study.
\subsection{Method} \label{subsec:external_listener_evaluation}

One of the authors, with tertiary qualifications and professional experience as an improvising saxophonist, participated in six improvisation sessions with the improvising machine to produce sixteen tracks.  Although the use of an author-participant has the potential to introduce unintended bias, they had greater experience playing with the system over participants used in the performer study and their improvisation sessions were selected to best highlight the effect of the differing communication conditions.  Prior research suggests that layperson comparative evaluations of computer generated music are sharpest once the musical output has reached a reasonable level of mainstream musical plausibility \cite{stowell2009evaluation,jordanous2013evaluating}.  

Randomised conditions and sampling were used to reduce any possible effects of unintended bias.  Truthful and deceptive visualisation conditions were alternated between randomly throughout the improvisation sessions, such that the participant did not know which condition was in use.  Six tracks were selected from the sixteen recordings by random stratified-sampling to balance the number of Truthful and Deceptive conditions and used for the listener study.  The first minute of each of these tracks was paired into a series of A/B comparisons that were embedded into a web questionnaire.  Audio files used in the questionnaire can be heard at \url{https://dx.doi.org/10.6084/m9.figshare.7552235.v1}

We used the Amazon Mechanical Turk platform to recruit 100 participants, who were asked to answer two questions for each of three A/B comparisons: `Which performance was more interesting?' and `Which performance had a better musical balance between drums and saxophone?'.  Each participant was given \$1 USD for participating.

Participants were not questioned on their musical interest or ability, but questions were included in the survey to identify participants who may have not understood the questions.  An additional A/B comparison was added with one track not containing drums.  This comparison also had an additional question asking which of the two recordings contained drums.  We excluded participants who found a better balance in the track with only saxophone, stated the wrong track contained drums was excluded, or spent less than the 8 minutes required to listen to all tracks on the questionnaire.  Of 100 participants, 96 met these requirements and their feedback was used for evaluation.

\subsection{Results and Discussion}
The results indicated a modest but significant tendency for the music produced in the Truthful visualisation condition to be perceived as more musically balanced than the Deceptive condition. Significance here is to be understood with respect to the number of participants (not the number of musical examples). A null hypothesis of random choice for the more musically balanced track is rejected by a test against the binomial distribution at the 95\% for these particular musical examples. This suggests a noticeable effect for external listeners.

With on-line listener studies there are a number of environmental conditions that are not controlled for, that can have a significant influence on the experience of listening to music.  Volume, speaker quality, speaker type and background noise can vary for each participant.  By framing the questions as a series of A/B comparisons, most of these factors would likely stay consistent for compared tracks.

Because live performance is a critical part of demonstrating improvisation, future evaluation of the machine improviser and effects of extra-musical communication is intended within live performance contexts.

\begin{table}[t]
\caption{Results of each A/B comparison in the on-line listener evaluation questionnaire.}
\label{tab:turkraw} 

\begin{tabular}{@{}ccc@{}}
\toprule
       & \multicolumn{2}{c}{Truthful Condition}    \\ \midrule
Tracks & More interesting & Better musical balance \\ \midrule
A \textit{vs.} B    & 44\%           & 51\%                 \\
C \textit{vs.} D    & 67\%           & 65\%                 \\
E \textit{vs.} F    & 57\%           & 60\%                 \\ 
\midrule
\textbf{Total }    & \textbf{53\%}	&  \textbf{55\%*}
\\
\bottomrule
\end{tabular}
\end{table}

\section{Conclusion}
As interaction with Artificial Intelligences -- and in particular creative improvisers -- becomes more commonplace, \textit{how} we interact and collaborate with co-creational AI systems is an increasingly important area of research. In this paper we have investigated how extra-musical cues between human and AI improvisers can affect the achievement of flow states in an improvising duet. Our results demonstrate that, at least for the performers evaluated, communication of a confidence metric improves the human performer's ability to achieve flow states more readily. Additionally, we demonstrated that the resulting improvisational performances are more likely to be perceived positively by non-expert audiences in terms of musical balance between instruements.

These results support our hypothesis that the extra-musical communication of `internal state' \textbf{\textit{of} the machine \textit{to} the human} can facilitate more engaging human-machine musical interactions.

Our experimentation with the use of biometric data (SC) as a proxy for the performer's musical engagement that could be communicated to the AI did not improve the performer's experience of musical flow, despite prior research showing this to be the case with pairs of improvising human musicians \cite{dean2015sc}. A difference between this previous study and our own was the location of the sensor: the Empatica E4 watch used in our experiments can only be worn on the wrist, whereas the previous study used a sensor attached to the ankle of the performer, to minimise inconsistencies due to movement which are common when improvising during performance. 

That extra-musical communication of `internal state'\textbf{\textit{of} the human musician \textit{to} the machine} would also facilitate more engaging human-machine musical interactions was not supported in this study but is also not ruled-out.  In further research we aim to draw on other kinds of biometric sensors that are more resilient to the effects of sudden moment that is a necessary part of physical playing and explore other modes of extra-musical communication.

Conceptualising intelligent machines as creative partners rather than passive tools or instruments is relatively new. While we have a rich and well explored history of improvisation between human performers to draw upon, improvising with an alien, non-human yet active participant creates many exciting new possibilities for human-machine partnerships. The challenge, which we have begun to explore in this paper, is to maximise the creative benefits and possibilities for both performers and audiences.

\bibliographystyle{ACM-Reference-Format}


\end{document}